\documentclass[preprint,prc,letterpaper,tightenlines,nofootinbib,showpacs,superscriptaddress]{revtex4}
\usepackage{graphicx}
\usepackage{amssymb,amsmath,mathrsfs}
\usepackage{dcolumn}
\usepackage{bm}

\newcommand{\nn}{\nonumber\\}

\newcommand{\uu}{\mathscr{U}}
\newcommand{\vv}{\mathscr{V}}
\newcommand{\la}{\langle}
\newcommand{\ra}{\rangle}
\newcommand{\bn}{\bar{n}}

\newcommand{\bPsi}{\Psi^\dagger}

\newcommand{\ben}{\begin{displaymath}}
\newcommand{\een}{\end{displaymath}}
\newcommand{\be}{\begin{equation}}
\newcommand{\ee}{\end{equation}}
\newcommand{\bea}{\begin{eqnarray}}
\newcommand{\eea}{\end{eqnarray}}

\newcommand{\ud}{\uu^\dagger}
\newcommand{\vd}{\vv^\dagger}

\newcommand{\hX}{\hat{X}}

\newcommand{\w}{\omega}

\newcommand{\tG}{\tilde{G}}
\newcommand{\tD}{\tilde{D}}
\newcommand{\tK}{\tilde{K}}
\newcommand{\tV}{\tilde{V}}
\newcommand{\tv}{\tilde{v}}

\newcommand{\ttG}{\langle {\cal G}\rangle}
\newcommand{\tr}{\mbox{Tr}}
\newcommand{\dPsi}{\Psi^\dagger}
\newcommand{\x}{{\bf x}}
\newcommand{\y}{{\bf y}}
\newcommand{\p}{{\bf p}}

\newcommand{\TT}{{\cal T}}

\newcommand{\tT}{\tilde{T}}

\newcommand{\bc}{\begin{center}}
\newcommand{\ec}{\end{center}}

\newcommand{\mat}[4]{\left( \begin{array}{cc} {#1} & {#2}  \\  {#3} &  {#4} \end{array} \right)}
\newcommand{\eqn}[1]{\label{#1}}
\newcommand{\eq}[1]{Eq.~(\ref{#1})}
\newcommand{\eqs}[1]{Eqs.~(\ref{#1})}

\begin{document}
\title{Three-body problem at finite temperature and density}
\author{A. N.  Kvinikhidze}
\affiliation{A.\ Razmadze Mathematical Institute, Georgian Academy of Sciences, Aleksidze Str.1, Tbilisi 0193, Georgia}
\affiliation{Department of Physics, Flinders University, Bedford Park, SA 5042, Australia}
\author{B. Blankleider}
\affiliation{Department of Physics, Flinders University, Bedford Park, SA 5042, Australia}
\date{\today}
\begin{abstract}
We derive practical three-body equations for the equal-time three-body Green function in 
matter.  Our equations describe both bosons and fermions at finite density and temperature, and take into account all possible two-body sub-processes allowed by the underlying Hamiltonian. 
\end{abstract}
\maketitle
\section{Introduction} 

Three-body correlations play an important role in describing many aspects of many-body systems.
In early studies of nuclear matter, where the main focus of interest was its binding energy,  three-body correlations were found to contribute a small but significant part \cite{Bethe,Rajaraman,Day1,Day2}. More recently, it has been the study of in-matter three-body systems themselves, that has been of main interest. Indeed, the in-matter three-body problem plays an important role in describing a large variety of interesting phenomena in many-body systems. For example, in order to understand the formation of bound states in heavy ion collisions, three-body calculations are needed to study the modification of
the binding energy and wave function of a three-nucleon bound state due to nuclear matter
of finite density and temperature \cite{Beyer1,Beyer2}. Similarly, studies of the binding energy of three quarks are of relevance to the understanding of color superconductivity and phase transitions in quark matter \cite{Beyer_quark,Mattiello}.
Three-body  calculations  are also needed to describe nonequilibrium processes of cluster formation in an interacting many-body system \cite{Beyer3,Beyer4}, and play a fundamental role in determining the two-particle-one-hole (pph) and two-hole-one-particle (hhp) contributions to the self-energy of the single-particle propagator \cite{Barbieri}.

The goal of the present paper is to formulate three-dimensional equations for the finite temperature in-matter three-body problem, that take into account all possible two-body sub-processes allowed by the underlying Hamiltonian. To put this goal into context, it is worthwhile to briefly review the progress made so far on this subject.  From the very beginning, it was recognized that Faddeev's approach \cite{Faddeev} provided a powerful tool in the description of few-body properties in Quantum Mechanics. It is therefore natural that not long after its formulation, this approach was also applied to quantum field theory, first within a four-dimensional formulation \cite{StoTav}, and then within a three-dimensional one obtained by equating times in four-dimensional Green functions \cite{kvst}. These early formulations were for three particles in vacuum. With the application of quantum field theoretical methods to statistical physics \cite{Abrikosov,walecka}, it became possible to apply Fadeev's approach also to the field-theoretic description of three particles within a many-body environment.
However, one major obstacle in formulating a practical description in this way, is the hole contribution to the single-particle propagator in the form  of an advanced part [see, for example, \eq{df}], which is not present in the quantum mechanical description of particles in vacuum. The presence of this hole contribution makes the field-theoretic description inherently four dimensional, even in the non-relativistic case. The first steps in applying the Faddeev approach to the many-body enivornment avoided this problem by utilizing a Bethe-Goldstone type of modification of the Faddeev equations, which involves of a simple momentum cut-off restricting the intermediate state particles to be above the Fermi surface \cite{Bethe}. Although such modified equations can be treated with the Faddeev method, they do not take into account the hole contributions which reside in the advanced parts of the single particle propagators. The way beyond this approximation was proposed by Schuck, Villars, and Ring \cite{schuck},  who used equal-time Green functions in order to obtain a three-dimensional field theoretic description. To derive their equation for the zero-temperature equal time three-body wave function, they approximated the effective pair-interaction kernels by terms linear in the physical two-body potentials. 
Since the exact expression for the effective pair-interaction kernel involves an infinite series of higher order terms as well [see \eq{tVipert}], the linear approximation cannot be considered as satisfactory for the strong coupling case, e.g., when two-body bound states are possible. The current state-of-the art formulation \cite{BR}, which has been used extensively for calculations \cite{Beyer1,Beyer2,Beyer_quark,Mattiello,Beyer3,Beyer4,BR} can be considered as the model of Ref.\ \cite{schuck} extended to finite temperatures, with the extension being performed using the imaginary time formalism of perturbation theory \cite{walecka}.

In this context, the goal of the present paper is to formulate practical field-theoretic  three-body equations, valid at finite temperature and density, that take into account the {\em whole} of the above mentioned series for the effective pair interaction. Like Refs.\ \cite{schuck} and \cite{BR}, we use equal-time Green functions to formulate a three-dimensional field theoretic description. We show that Faddeev's idea which renders the three-body kernel compact, namely, to reexpress the three-body equations in terms of two-body t matrices rather than two-body potentials, also enables one to sum up, {\em exactly}, the infinite series of \eq{tVipert} for the pair-interaction kernel.

\section{In-matter four-dimensional three-body equations}

The interactions of three identical particles at finite density and temperature
are described in quantum field theory by the Green function ${\cal G}$ defined by
\begin{eqnarray}\label{G6pt}
\lefteqn{(2\pi)^4\delta^4(p'_1+p'_2+p'_3-p_1-p_2-p_3){\cal G}(p'_1p'_2p'_3;p_1p_2p_3)= 
\int d^4y_1d^4y_2d^4y_3d^4x_1d^4x_2d^4x_3} 
\nn
&&e^{i(p'_1\cdot y_1+p'_2\cdot y_2+p'_3\cdot y_3-p_1\cdot x_1-p_2\cdot x_2-p_3\cdot x_3)}\,
\mbox{Tr}\left\{ \rho\, \TT [\Psi(y_1)\Psi(y_2)\Psi(y_3)
\bPsi(x_3)\bPsi(x_2)\bPsi(x_1)]\right\}
\end{eqnarray}
where $\Psi$ and $\bPsi$ are Heisenberg fields with respect to the Hamiltonian $K=H-\mu N$,
$\TT$ is the time ordering operator and
\be
\rho = \frac{e^{-\beta K} }{\mbox{Tr}\, e^{-\beta K} }
\ee
is the statistical operator of the grand canonical ensemble \cite{walecka}. Besides being the central quantity for the description of three-body observables, this Green function is also needed
to calculate the vacuum properties of the system with the help of the dressed
single particle propagator; for example, in the four-point interaction model,
the single particle self-energy diagram is completely defined by the particle-particle-hole (pph)
Green function \cite{dukelsky}.

In the zero temperature case ($\beta=1/k_BT\rightarrow \infty$) only the ground state survives in the trace of \eq{G6pt}, and 
the Green function reduces to the usual QFT expectation value
\begin{eqnarray}
\lefteqn{(2\pi)^4\delta^4(p'_1+p'_2+p'_3-p_1-p_2-p_3){\cal G}(p'_1p'_2p'_3;p_1p_2p_3)= 
\int d^4y_1d^4y_2d^4y_3d^4x_1d^4x_2d^4x_3} 
\hspace{1.0cm}\nn
&&e^{i(p'_1\cdot y_1+p'_2\cdot y_2+p'_3\cdot y_3-p_1\cdot x_1-p_2\cdot x_2-p_3\cdot x_3)}\,
\langle 0|\TT\Psi(y_1)\Psi(y_2)\Psi(y_3)
\bPsi(x_3)\bPsi(x_2)\bPsi(x_1)|0\rangle
\end{eqnarray}
where $|0\rangle$ is the physical ground state. For the latter, straightforward
use of Wick's theorem gives a perturbation theory with Feynman rules.
Analogously, two types of perturbation theory, so-called "imaginary-time"  and "real-time",
have been derived for \eq{G6pt} \cite{LeBellac}.

Here we shall use the real-time formulation of perturbation theory in which the number of degrees of freedom is doubled \cite{smilga}; this complication, with respect to the zero-temperature case, comes from the sum over the complete set of states  (trace) in \eq{G6pt}.\footnote{A similar discussion on the basis of the imaginary-time formalism and its comparison with this note will be presented elsewhere.}
For example, the free one-body propagator, in the nonrelativistic case, is given by \cite{umezawa}
\be\label{df}
d^f(p)=i\left[\frac{\bn(\p)}{p^0-\w+ i \epsilon} +
\frac{n(\p)}{p^0-\w- i \epsilon} \right]
\ee
where $\w=\w_\p=\p^2/2m - \mu$ (we take $\hbar=1$), $\mu$ is the chemical potential, and $n, \bar n$ are $2\times 2$ matrices whose elements are simple functions of the distribution function
\be
f(\w) = \frac{1}{e^{\beta\w}\pm 1},
\ee
with the upper sign ($+$) for fermions and the lower sign ($-$) for bosons (see Appendix A). Correspondingly, the elementary vertices have an extra double-valued index for 
each particle leg; e.g., the four-point interaction $\bar{v}$ used to define the
Hamiltonian of \eq{Ham}, enters the formalism with doubled degrees of freedom through the quantity $\bar{\mathrm{v}}$ whose matrix structure is given as
$\bar{\mathrm{v}}_{ijkl}=(\delta_{i1}\delta_{j1}\delta_{k1}\delta_{l1}-
\delta_{i2}\delta_{j2}\delta_{k2}\delta_{l2})\bar v$. Note that the first diagonal element is just the potential itself, $\bar{\mathrm{v}}_{1111}=\bar{v}$. Similarly, the first diagonal element of \eq{df}, $d^f_{11}$, for which 
\begin{equation}\label{simpl} 
n_{11}(\p)=1- \bn_{11}(\p)= \pm \frac{1}{e^{\beta \w}\pm1} = \pm f(\w),
\ee
corresponds to the usual one-body free Green function at zero temperature \cite{walecka}.
In some cases one can use a simplified propagator consisting of just  this
$d^f_{11}$ element of \eq{df} \cite{jackiw}.

For the identical particle case considered here,  the field theoretic expression of \eq{G6pt} automatically guarantees the appropriate symmetry of the three-particle Green function ${\cal G}$.  Moreover, in the doubled degrees of freedom formalism, the matrix Green function $G$ whose first diagonal element is ${\cal G}$, is likewise properly symmetric in the case of bosons, and antisymmetric in the case of fermions. On the other hand, the disconnected Green function $G_0$ defined by
\be
G_0(p_1' p_2' p_3',p_1p_2p_3)=d(p_1)\, d(p_2)\, d(p_3)\, (2\pi)^4 \delta^4(p_2'-p_2)\,
(2\pi)^4\delta^4(p_3'-p_3) ,   \eqn{G_0123}
\ee
where $d(p_i)$ is the dressed propagator of particle $i$, does not possess identical particle symmetry; thus $G_0$ is not equal to the
fully disconnected part of $G$ (which we shall denote by $G_d$).  Indeed, it can
be easily shown  that to obtain $G_d$, one need only symmetrize (or antisymmetrize) $G_0$ according to the equation
\be
\sum_P G_0(1'2'3',123) \equiv G^P_0(1'2'3',123) = G_d(1'2'3',123)     \eqn{calG_0}
\ee
where the sum is over all permutations $P$ of either the initial or final state
particle labels, and for fermions, is understood to include a factor $(-1)^P = +1$ or $-1$
depending on whether the permutation is even or odd, respectively. In
\eq{calG_0} we use a symbolic notation where integers represent the momenta plus
all quantum numbers of the corresponding particles, with primes distinguishing
the final states.

Defining the kernel $K$ to be the set of all possible three-particle irreducible
Feynman diagrams for the $3\rightarrow 3$ process, we may write the
equation for the Green function $G$ as \cite{kbgaug}
\begin{equation}\label{G_as}
G = G_0^P + \frac{1}{3!} G_0 K G       
\end{equation}
where the $1/3!$ factor reflects the fact that both $G$ and $K$ are fully
symmetric or antisymmetric in their particle labels. 
The disconnected part of $K$, indicated by subscript $d$, can be expressed in terms of the identical particle two-body potential $v$ as
\be
K_d(1'2'3',123)=\sum_{L_cR_c} v(2'3',23) d^{-1}(1)\delta(1',1)
\ee
where $\delta(1',1)$ represents the momentum conserving Dirac $\delta$ function
$(2\pi)^4\delta^4(p_1'-p_1)$, while $L_c$ and $R_c$ indicate that sums
are taken over cyclic permutations of the left labels $(1'2'3')$ and right labels
$(123)$, respectively (note that the sums are restricted to cyclic permutations
because the potential $v$ is already properly symmetric or antisymmetric in its labels).

Defining
\be
V_i(1'2'3,123) = v(j'k',jk) d^{-1}(i)\delta(i',i)      \eqn{V_i_sym}
\ee
where $(ijk)$ is a cyclic permutation of $(123)$, we have that
\be
K_d = \sum_{P_c} (V_1+V_2+V_3),
\ee
where it makes no difference over which labels, left or right, the cyclic
permutations are taken. 
Denoting the connected part of the kernel by $K_c$, we define the $3\rightarrow
3$ potential $V$ by
\be
V=\frac{1}{2}(V_1+V_2+V_3) + \frac{1}{6}K_c.   \eqn{V_sym}
\ee
Although $V$ is not fully symmetric or antisymmetric, it does have the useful symmetry property
\be
P_{ij} V P_{ij} = V    \eqn{V-ij-symm}
\ee
where $P_{ij}$ is the operator that exchanges the $i$'th and $j$'th momentum,
spin, and isospin labels. Since
\be
K=\sum_P  V,
\ee
\eq{G_as} can be written as
\be
G = G_0^P + G_0 V G .      \eqn{Gas}
\ee
Formally, \eq{Gas} differs from the equivalent relation for distinguishable
particles, $G = G_0 + G_0 V G$, only in the explicit symmetrization 
of the inhomogeneous term. 

For the sake of simplicity, we consider the model given by the second quantized Hamiltonian
\be
H=\sum_1\w_1a^\dagger_1a_1+\sum_{1234}\bar 
v(1234)a^\dagger_1a^\dagger_2a_4a_3 \eqn{Ham}
\ee
where $\w_1={\bf q}_1^2/2m$ is the single-particle non-relativistic kinetic energy, and $\bar v(1234)=\bar v(s_1 {\bf q}_1,s_2 {\bf q}_2, s_3 {\bf q}_3, s_4 {\bf q}_4)$ is a Gallilei invariant function of 
spins $s_i$ and three-dimensional momenta ${\bf q}_i$ of 
particles $1, 2, 3$ and $4$. 
Then the first diagonal element of the pair interaction potential $v$ of \eq{V_i_sym}, $v_{1111}$, is not $\bar v$ in the general case of finite density and temperature, as the presence of the medium
makes it possible for particle-hole pairs to be exchanged in the t-channel. As a result
of these exchanges $v=v(s_1 q_1,s_2 q_2, s_3 q_3, s_4 q_4)$ also depends on the 
energy variables, $q^0_1, q^0_2, q^0_3, q^0_4$ (of which only three are independent since
$q^0_1+q^0_2=q^0_3+q^0_4$). In the case of
zero density and temperature, $\mu=0, T=0$, we obtain pure quantum mechanics 
with particle number conservation and $v_{1111}=\bar v$. In this case, single particle 
propagators do not have an advanced part, $n=0$, and this fact allows one to derive the standard
three-dimensional Faddeev equations for the three-particle system where input two-body t matrices (in three-body space) are obtained from physical two-body t matrices (in two-body space) simply by subtracting the spectator energy from the total energy. In the next two sections we discuss how three-dimensional three-body equations can also be derived for the case of finite temperature and density.

\section{Three-dimensional equal-time reduction}

Being a four-dimensional integral equation, \eq{Gas} is not very convenient for practical calculations -  it involves integrations over relative times (or relative energies) which, because of the presence of cuts and singularities in all four quadrants of the integration plane, cannot be easily handled numerically.
For this reason, we would like to implement a three-dimensional reduction of this equation. To do this we follow the current literature and effect this reduction by equating times in initial states, and separately, in final states. A quantity $A$ with such equated times will be denoted by $i\la A\ra$ (an extra factor of $i$ has been included for later convenience). Thus, our central quantity is the two-time Green function $\ttG$ which is obtained from the four-dimensional Green function ${\cal G}$ by equating times as just described. In momentum space, $\ttG$ is therefore given by
\begin{align}
(2&\pi)^3\delta^3(\p'_1+\p'_2+\p'_3-\p_1-\p_2-\p_3)\, i\ttG(E,\p'_1\p'_2\p'_3,\p_1\p_2\p_3)\nn[2mm]
&=\int d^3y_1\,d^3y_2\,d^3y_3\,d^3x_1\,d^3x_2\,d^3x_3\, dt\,
e^{i(Et-\p'_1\cdot \y_1-\p'_2\cdot \y_2-\p'_3\cdot \y_3+\p_1\cdot \x_1+\p_2\cdot \x_2+\p_3\cdot \x_3)}\nn[2mm]
&\hspace{1cm}\tr \left\{ \rho {\cal T} [\Psi(t,\y_1)\Psi(t,\y_2)\Psi(t,\y_3)\dPsi(0,\x_3)\dPsi(0,\x_2)\dPsi(0,\x_1)]\right\} .   \eqn{etG6pt}
\end{align}
In the doubled degrees formalism, the two-time Green function of \eq{etG6pt} is the first diagonal element of the matrix Green function $\la G\ra$. The goal of this section is to develop the three-dimensional integral equation, analogous to \eq{Gas}, for $\la G\ra$.
Previously, such three-dimensional equations for the two-time Green function have been considered in the context of (zero density) relativistic quantum field theory in Ref.\ \cite{log} for the two-particle case, and in Ref.\  \cite{kvst} for the three-particle case. We shall base our derivation of the in-matter three-body equation upon that of Ref.\ \cite{kvst}. By contrast, other recent many-body formulations of three-body equations have been
closely related to the work of Ref.\ \cite{schuck}. The difference between these two approaches will be discussed in due course.

It can be shown that the momentum space two-time Green function of \eq{etG6pt} can be obtained directly from the four-dimensional Green function of \eq{G6pt} by integrating out all the relative energies:
\begin{align}
i \ttG (E, \p'_1\p'_2 & \p'_3,\p_1\p_2\p_3)
=\int  \frac{d{p'_1}^{0}}{2\pi} \,\frac{d{p'_2}^{0}}{2\pi}\,\frac{d{p'_3}^{0}}{2\pi}
\,\frac{dp_1^{0}}{2\pi}\,\frac{dp_2^{0}}{2\pi}\,\frac{dp_3^{0}}{2\pi} \,  {\cal G}(p'_1p'_2p'_3,p_1p_2p_3)\nn[1mm]
&(2\pi)^2\, \delta({p'_1}^{0}+{p'_2}^{0}+{p'_3}^{0}-E)\, \delta(p_1^{0}+p_2^{0}+p_3^{0}-E).
\end{align}
Correspondingly, in the doubled degrees of freedom formalism, the equal time matrix Green function $\la G\ra$ is related to its four-dimensional counterpart $G$ by
\begin{align}
i \la G \ra (E, \p'_1\p'_2 & \p'_3,\p_1\p_2\p_3)
=\int  \frac{d{p'_1}^{0}}{2\pi} \,\frac{d{p'_2}^{0}}{2\pi}\,\frac{d{p'_3}^{0}}{2\pi}
\,\frac{dp_1^{0}}{2\pi}\,\frac{dp_2^{0}}{2\pi}\,\frac{dp_3^{0}}{2\pi} \, G(p'_1p'_2p'_3,p_1p_2p_3)\nn[1mm]
&(2\pi)^2\, \delta({p'_1}^{0}+{p'_2}^{0}+{p'_3}^{0}-E)\, \delta(p_1^{0}+p_2^{0}+p_3^{0}-E).
\eqn{tt_main}
\end{align}
Thus the four-dimensional free three-body Green function, defined as
\be
G^f_0(p_1' p_2' p_3',p_1p_2p_3)=d^f(p_1)\, d^f(p_2)\, d^f(p_3)\, (2\pi)^8 \delta^4(p_2'-p_2)\,
\delta^4(p_3'-p_3) ,   \eqn{Gf_0123}
\ee
where $d^f$ is given by \eq{df}, leads to the following two-time free three-body Green function:
\be
\la G^f_0\ra(E,\p_1' \p_2' \p_3',\p_1\p_2\p_3)=\la G_0^f\ra(E,\p_1\p_2\p_3)\, (2\pi)^6 \delta^3(\p_2'-\p_2)\,
\delta^3(\p_3'-\p_3) , 
\ee
where\footnote{To save on notation, we use the same symbol with differing numbers of arguments to represent disconnected Green functions with and without the momentum conserving $\delta$ functions.}
\begin{align}
\la G_0^f\ra (E,\p_1&\p_2\p_3)
=-i \int \frac{dp_2^{0}}{2\pi}\,\frac{dp_3^{0}}{2\pi} \, 
d^f(E-p_2^0-p_3^0,\p_1)\,  d^f(p_2^0,\p_2)\, d^f(p_3^0,\p_3)\nn[2mm]
&=\frac{\bn(\p_1)\bn(\p_2)\bn(\p_3)}
{E-\w_{\p_1}-\w_{\p_2}-\w_{\p_3}+ i \epsilon}+
\frac{n(\p_1) n(\p_2) n(\p_3)}
{E-\w_{\p_1}-\w_{\p_2}-\w_{\p_3}- i \epsilon} .   \eqn{G0f}
\end{align} 

In a similar way, we can define the Green function ${\cal D}$ of two identical particles at finite density and temperature as
\begin{align}\label{G4pt}
(2\pi)^4\delta^4(p'_1+p'_2-p_1&-p_2){\cal D}(p'_1p'_2;p_1p_2)= 
\int d^4y_1d^4y_2 d^4x_1d^4x_2 \nn
&e^{i(p'_1\cdot y_1+p'_2\cdot y_2-p_1\cdot x_1-p_2\cdot x_2)}\,
\mbox{Tr}\left\{ \rho\, \TT [\Psi(y_1)\Psi(y_2)\bPsi(x_2)\bPsi(x_1)]\right\}
\end{align}
from which the two-time Green function $\la {\cal D}\ra$ follows:
\begin{align}
i  \la {\cal D}\ra(E, \p'_1\p'_2,\p_1\p_2)
=\int  \frac{d{p'_1}^{0}}{2\pi}& \,\frac{d{p'_2}^{0}}{2\pi}\,
\,\frac{dp_1^{0}}{2\pi}\,\frac{dp_2^{0}}{2\pi}\, {\cal D}(p'_1p'_2,p_1p_2)\nn[1mm]
&(2\pi)^2\, \delta({p'_1}^{0}+{p'_2}^{0}-E)\, \delta(p_1^{0}+p_2^{0}-E).
\end{align}
In the doubled degrees of freedom formalism a similar equation relates the two-body matrix Green function $D$ to its two-time version $\la D\ra$. Applying this relation to the free two-body Green function
\be
D^f_0(p_1' p_2',p_1p_2)=d^f(p_1)\, d^f(p_2)\, (2\pi)^4 \delta^4(p_2'-p_2),   \eqn{D_012}
\ee
one obtains the two-time free two-body Green function:
\be
\la D^f_0\ra(E,\p_1' \p_2',\p_1\p_2)=\la D_0^f\ra(E,\p_1\p_2)\, (2\pi)^3 \delta^3(\p_2'-\p_2), 
\ee
where
\begin{align}
\la D_0^f\ra(E,\p_1\p_2)
&=-i \int \frac{dp_2^{0}}{2\pi}\,
d^f(E-p_2^0,\p_1)\,  d^f(p_2^0,\p_2)  \nn[2mm]
&=\frac{\bn(\p_1)\bn(\p_2)}
{E-\w_{\p_1}-\w_{\p_2}+ i \epsilon}-
\frac{n(\p_1) n(\p_2)}
{E-\w_{\p_1}-\w_{\p_2}- i \epsilon} .   \eqn{D0f}
\end{align} 
It also convenient to express the one-body propagator as
\be
\la d^f\ra(E,\p) = \left. -i d^f(p)\right|_{p^0=E}= \frac{\bn(\p)}{E-\w_\p+ i \epsilon} +
\frac{n(\p)}{E-\w_\p- i \epsilon}.   \eqn{df_tt}
\ee

\subsection{Three-body equal time quasi-potential}

One can write a three-dimensional equation for the two-time Green function $\la G\ra$ of the same form as four-dimensional \eq{Gas}:
\be
\la G \ra = \la G_0\ra^P + \la G_0\ra \tV \la G\ra,    \eqn{Gas3d}
\ee
where $\tV$ is the three-dimensional "quasi-potential". As is well known \cite{log}, $\tV$ should be expressible in terms of a perturbation series in the four-dimensional potential $V$; however, this series involves the inverse of the disconnected Green function $\la G_0\ra$, which in the many-body case is a problem, as this inverse may not exist. To see this, it is sufficient to consider the free propagator of \eq{G0f} for the case of fermions at $T=0$ where real time perturbation theory is expressed in single degrees of freedom. In this case the functions $n$ and $\bn$ reduce to
\be
n(\p) = \theta(p_F-|\p|),\hspace{1cm} \bn(\p) = 1 - n(\p) = \theta(|\p|-p_F)
\ee
where $p_F=\mu(0)$ is the Fermi momentum. It is clear that $n$ and $\bn$ are projection operators, and this in turn means that $\la G_0^f\ra$ is also a projection operator. Thus  $\la G_0^f\ra$ does not project onto the full space of momenta, but only onto the sub-space projected by the operator
\be
{\cal N} = n(\p_1)n(\p_2)n(\p_3) +  \bn(\p_1)\bn(\p_2)\bn(\p_3);
\ee
as a result, $\la G_0^f\ra$ cannot be inverted in the full space of momenta. Exactly the same conclusion is reached in the case of finite temperature in the formalism with doubled degrees of freedom, as $n$ and $\bn$ remain projection operators, but now as matrices. More specifically (see Appendix A), $n$ and $\bn$ satisfy the following relations which define their projection properties: $n + \bn = g$, 
$n\, g\, \bn = \bn\, g\, n = 0$,  $n \,g \,n = n$, and  $\bn\, g\,\bn = \bn$ where $g$ is defined as 
\be
g=\mat{1}{0}{0}{1} \ \mbox{for fermions},\hspace{1cm} g=\mat{1}{0}{0}{-1} \ \mbox{for bosons}.
\ee
To get around the non-invertability of $\la G_0\ra $, we introduce a modified Green function
\be
\tG_0 = \la G_0\ra + (ggg-{\cal N})\Delta   \eqn{tG0}
\ee
which for non-zero $\Delta$ is not singular and can be inverted. In \eq{tG0}, $ggg$ is a direct product of $g$'s where one $g$ is in the space of particle 1, another is in the space of particle 2, and the third is in the space of particle 3.
It is important to note that $\tG_0$ is identical to $\la G_0\ra$ in the subspace projected by ${\cal N}$, i.e.,  ${\cal N}ggg\tG_0 = {\cal N}ggg\la G_0\ra = \la G_0\ra$.
The operator $\Delta$ is required to be fully disconnected, so that
\be
\tG_0(E,\p'_1\p'_2\p'_3,\p_1\p_2\p_3) = \tG_0(E,\p_1\p_2\p_3)\, (2\pi)^6 \delta^3(\p'_2-\p_2)\,\delta^3(\p'_3-\p_3),
\ee
but can otherwise be chosen according to one's own convenience (later we will also be able to let $\Delta$ go safely to zero). For the free case, one can write down the inverse of $\tG^f_0$ explicitly:
\be
(\tG_0^f)^{-1}(E,\p_1\p_2\p_3) = ggg\left[ {\cal N}(E-\omega_{\p_1}-\omega_{\p_2}-\omega_{\p_3}) + (ggg-{\cal N})\Delta^{-1}\right]ggg. \eqn{tG0_inv}
\ee

As a consequence of introducing $\tG_0$, we shall likewise introduce a modified full three-dimensional Green function $\tG$ defined as
\be
\tG= \la G\ra + (ggg-{\cal N})\Delta^P   \eqn{tG}
\ee
and redefine the three-dimensional quasi-potential $\tV$ to satisfy the equation
\be
\tG = \tG_0^P + \tG_0\tV \tG.   \eqn{Gasn}
\ee

As it stands, \eq{Gas} is not convenient to work with as the inhomogeneous term contains an explicit sum over the permutations of $G_0$'s labels. For this reason we define the unsymmetrized Green function  $G^u$ as the solution of
\be
G^u = G_0 + G_0 V G^u.    \eqn{G^u}
\ee
The full Green function $G$ is then obtained by summing over the permutations of the right-hand labels of $G^u$, which we symbolically write as
$
G=G^u P.
$
Similarly, in the three-dimensional case we define
\be
\tG^u = \tG_0 + \tG_0\tV \tG^u   \eqn{tV1}
\ee
where $\tG=\tG^u P$. It follows  that
\be
\tG^u= \la G^u\ra + (ggg-{\cal N})\Delta.    \eqn{tGu}
\ee
Iterating \eq{G^u}, equating initial and final times, and using \eqs{tG0} and (\ref{tGu}), we obtain
\be
\tG^u = \tG_0 + \la G_0 V G_0\ra+ \la G_0 V G_0V G_0\ra + \ldots \eqn{pert}
\ee
where in momentum space the angle brackets indicate the integration of relative energies as in \eq{tt_main}. 
Since the inverse $\tG_0^{-1}$ exists by construction,
\be
\tG_0^{-1}\tG^u = 1+ \tG_0^{-1}\la G_0 V G_0\ra+ \tG_0^{-1}\la G_0 V G_0V G_0\ra + \ldots
\ee
so that
\begin{align}
(\tG^u)^{-1}\tG_0 = 1 &- \tG_0^{-1}\la G_0 V G_0\ra- \tG_0^{-1}\la G_0 V G_0V G_0\ra\nn
& + \tG_0^{-1}\la G_0 V G_0\ra \tG_0^{-1}\la G_0 V G_0\ra -\ldots
\end{align}
Using this in \eq{tV1} gives an explicit perturbation series for the quasi-potential:
\be
\tV =  \tG_0^{-1}\left[ \la G_0 V G_0\ra + \la G_0 V G_0V G_0\ra
- \la G_0 V G_0\ra\tG_0^{-1}\la G_0 V G_0\ra + \ldots\right]  \tG_0^{-1} . \eqn{tVpert}
\ee
In the present case of a three-particle system, $V$ consists of a sum of pair interactions and three-body forces,  as given in \eq{V_sym}.  The quasi-potential $\tV$ must be expressible similarly as
\be
\tV=\frac{1}{2}\left(\tV_1+\tV_2+\tV_3\right) + \frac{1}{6}\tK_c  \eqn{tV}
\ee
where $\tV_i$ is a pair interaction with particle $i$ as spectator, and $\tK_c$ is a three-particle irreducible connected term (the three-body force). We can thus write $\tV_i$ as
\be
\tV_i(\p'_{1}\p'_{2}\p'_{3},\p_{1}\p_{2}\p_{3}) = (2\pi)^3\delta^3(\p'_i-\p_i) \tV_i(\p'_j\p'_k,\p_j\p_k;\p_i)   \eqn{tVi}
\ee
where, to save on notation, we have used the same symbol $\tV_i$, but with different arguments, to denote the pair interaction with and without the spectator $\delta$ function. Note that, in general, the $\tV_i$ without the $\delta$ function still depends on the momentum $\p_i$ of the spectator. It follows from \eq{tVpert} that  
\be
\frac{\tV_i}{2} =  \tG_0^{-1}\left[ \la G_0 \frac{V_i}{2} G_0\ra + \la G_0 \frac{V_i}{2} G_0\frac{V_i}{2} G_0\ra
- \la G_0 \frac{V_i}{2} G_0\ra\tG_0^{-1}\la G_0 \frac{V_i}{2} G_0\ra + \ldots\right]  \tG_0^{-1} . \eqn{tVipert}
\ee

We are interested in the usual case where the three-body force $\tK_c$ is neglected. Then $\tV_i$ provides the sole interaction that describes three-body observables. However,
what enters the three-body Faddeev equations is the pair-interaction t matrix in three-body space, $\tT_i$, defined in terms of the quasi-potential $\tV_i$ by the Lippmann-Schwinger equation
\be
\tT_i = \tV_i + \frac{1}{2}\tV_i \tG_0 \tT_i . \eqn{LS}
\ee
In correspondence with \eq{tVi}, we shall write the disconnectedness structure of $\tilde{T}_i$ as
\be
\tilde{T}_i(\p'_{1}\p'_{2}\p'_{3},\p_{1}\p_{2}\p_{3}) = (2\pi)^3\delta^3(\p'_i-\p_i) \tilde{T}_i(\p'_j\p'_k,\p_j\p_k;\p_i) .  \eqn{tTi}
\ee
The task of constructing $\tT_i$ appears to be formidable. Just to determine $\tV_i$ by summing the infinite series of \eq{tVipert} seems already a practical impossibility.  That is why most, if not all, works on this subject keep only the linear term in the input two-body interaction \cite{schuck}. In our case, this would mean keeping only the first term of the series in \eq{tVipert}. 
However, as we shall show subsequently, there is another way of solving this problem which gives the exact t matrix $\tT_i$, namely the one that results from a complete summation of \eq{tVipert} followed by an exact solution of the Lippmann-Schwinger equation, \eq{LS}.

We finish this subsection by pointing out that the pair interaction part of the three-body quasi-potential, $\tV_i$ (without the spectator $\delta$ function),  is related in a most non-trivial way to the corresponding two-body quasi-potential $\tv_i$ arising from equating times in the two-body Green function. Indeed if we write the two-body equivalent of \eq{Gas},
\be
D = D_0^P + \frac{1}{2} D_0 v D,
\ee
where $D$ is the two-body Green function originating from \eq{G4pt}, $D_0$ is the two-body disconnected Green function given by
\be
D_0(p'_1p'_2,p_1p_2) = d(p_1) d(p_2) (2\pi)^4 \delta^4(p'_2-p_2),
\ee
and $v$ is the fully symmetric (or antisymmetric) two-body kernel, then the two-body quasi-potential $\tv$ can be determined by following the same procedure as the one above to determine $\tV$. Clearly, we shall obtain the following perturbation series for $\tv$:
\be
\frac{\tv}{2} =  \tD_0^{-1}\left[ \la D_0 \frac{v}{2} D_0\ra + \la D_0 \frac{v}{2} D_0\frac{v}{2} D_0\ra
- \la D_0 \frac{v}{2} D_0\ra\tD_0^{-1}\la D_0 \frac{v}{2} D_0\ra + \ldots\right]  \tD_0^{-1}  \eqn{tvpert}
\ee
where
\be
\tD_0 = \la D_0\ra + (gg - {\cal N}^{(2)}) \Delta^{(2)}   \eqn{tD_0}
\ee
with
\be
{\cal N}^{(2)} = n(\p_1)n(\p_2)+\bn(\p_1)\bn(\p_2), \eqn{N2}
\ee
 and $\Delta^{(2)}$ is any convenient disconnected function that makes $\tD_0$ invertible. It is easy to see that in general, $\tV_i$ will be related to $\tv_i$ in a complicated and non-linear way through an infinite series.

\subsection{Exact three-body equal-time disconnected t matrix}

As the determination of the exact quasi-potential $\tV_i$ is difficult, as discussed above, it would seem that  the exact determination of the corresponding t matrix $\tT_i$, needed as input to the three-dimensional Faddeev equations, is not attainable, at least not through the iteration of the exact $\tV_i$. Yet it often happens that working with t matrices {\em directly} is much simpler than working with the underlying potential - one example being the problem of disconnectedness in the three-body problem, which Faddeev solved by formulating equations in terms of two-body t matrices rather than two-body potentials.

Thus, rather than expressing $\tV_i$ in terms of $V_i$, as we have done above in \eq{tVipert}, we shall now attempt to express the exact $\tT_i$ directly in terms of the corresponding four-dimensional disconnected t matrix $T_i$.
We start by noting that even though $\tV_i$ may not be known, one can nevertheless formally write down the exact $\tT_i$ as the solution of the Lippmann-Schwinger equation, \eq{LS}. Thus, if $\tG^u_i$ is the disconnected part of $\tG^u$ with particle $i$ as spectator, so that, by \eq{tV1},
\be
\tG^u_i = \tG_0 + \frac{1}{2} \tG_0 \tV_i \tG^u_i,
\ee
it follows that
\be
\tG^u_i = \tG_0 + \frac{1}{2} \tG_0 \tilde{T}_i \tG_0.  \eqn{tGi}
\ee
Defining the equal time three-body t matrix $\tilde{T}$ through
\be
\tG^u = \tG_0 + \tG_0 \tilde{T} \tG_0,  \eqn{tGuT}
\ee
it's clear that one can write
\be
\tilde{T} = \frac{1}{2}\left( \tilde{T}_1 +  \tilde{T}_2 +  \tilde{T}_3 \right) + \tilde{T}_c
\ee
where $\tilde{T}_c$ the part of $\tilde{T}$ which is fully connected.
 
In the four-dimensional sector one can similarly express the unsymmetrized Green function $G^u$ in terms of the three-body t matrix $T$ as
\be
G^u = G_0 + G_0 T G_0
\ee
where
\be
T  = V + V G_0 T .    \eqn{BS}
\ee 
 By analogy with \eq{V_i_sym} and \eq{V_sym} we write
 \be
 T = \frac{1}{2}(T_1+T_2+T_3) + T_c
 \ee
 where
 \be
 T_i(1'2'3',123) = t(j'k',jk) d^{-1}(i) \delta(i',i)   \eqn{T_i}
 \ee
 is the disconnnected part of $T$ with particle $i$ as spectator and $t(j'k',jk)\equiv t_i$ is the four-dimensional two-body t matrix, and $T_c$ is the connected part. It follows that
 \be
T_i  = V_i + \frac{1}{2}V_i G_0 T_i .    \eqn{BS_i}
\ee 
The disconnected part of $G^u$, with particle $i$ as spectator, is given by
\be
G_i^u = G_0 + \frac{1}{2}G_0 T_i G_0.   \eqn{Gi}
\ee
Performing the equal time operation on \eq{Gi},
\be
\la G_i^u\ra = \la G_0\ra + \frac{1}{2}\la G_0 T_i G_0\ra,   \eqn{Gi_eq}
\ee
it follows from \eq{tG0} and \eq{tGu} that
\be
\tG_i^u = \tG_0 + \frac{1}{2}\la G_0 T_i G_0\ra.   \eqn{tGi_eq}
\ee
Comparing this equation with \eq{tGi}, we obtain
\be
 \tG_0 \tT_i \tG_0 = \la G_0 T_i G_0\ra.   \eqn{main}
\ee
It is seen that in contrast to the quasi-potential $\tV_i$ which is related to the four-dimensional potential $V_i$ in a very complicated way, the t matrix $\tT_i$ corresponding to the quasi-potentional, defined by the {\em exact} solution of \eq{LS}, is connected to the four-dimensional t matrix $T_i$ in a very simple way.
 
 Indeed, using \eq{tt_main} and \eq{T_i}, one can write the RHS of  \eq{main} as
\begin{align}
\la G_0 T_i G_0 \ra &= -i
\int  \frac{d{p'_1}^{0}}{2\pi} \,\frac{d{p'_2}^{0}}{2\pi}\,\frac{d{p'_3}^{0}}{2\pi}
\,\frac{dp_1^{0}}{2\pi}\,\frac{dp_2^{0}}{2\pi}\,\frac{dp_3^{0}}{2\pi} \,  
[D_{0i} t_i D_{0i}](p'_jp'_k,p_jp_k) (2\pi)^4\delta^4(p'_i-p_i) d_i(p_i)\nn[1mm]
&(2\pi)^2\, \delta({p'_1}^{0}+{p'_2}^{0}+{p'_3}^{0}-E)\, \delta(p_1^{0}+p_2^{0}+p_3^{0}-E)\nn[5mm]
&= -i(2\pi)^3\delta^3(\p'_i-\p_i)
\int \frac{d{p'_j}^{0}}{2\pi}\,\frac{d{p'_k}^{0}}{2\pi}
\,\frac{dp_i^{0}}{2\pi}\,\frac{dp_j^{0}}{2\pi}\,\frac{dp_k^{0}}{2\pi} \,  
[D_{0i} t_i D_{0i}](p'_jp'_k,p_jp_k) d_i(p_i)\nn[1mm]
&(2\pi)^2\, \delta({p_i}^{0}+{p'_j}^{0}+{p'_k}^{0}-E)\, \delta(p_i^{0}+p_j^{0}+p_k^{0}-E)\nn[5mm]
&= (2\pi)^3\delta^3(\p'_i-\p_i)
\int \frac{dp_i^0}{2\pi} i
\la D_{0i} t_i D_{0i}\ra (E-p_i^0,\p'_j\p'_k,\p_j\p_k) \,\la d_i\ra (p_i^0,\p_i)
\end{align}
where
\begin{align}
[D_{0i}t_iD_{0i}]  (p'_jp'_k,p_jp_k) = 
d(p'_j)d(p'_k)\, t(p'_jp'_k,p_jp_k)\,d(p_j)d(p_k).
\end{align}
The result of \eq{main} can thus be written as
\be
\tG_0 \tT_i\tG_0 =   (2\pi)^3\delta^3(\p'_i-\p_i) \,  \la D_{0i} t_i D_{0i}\ra \otimes \la d_i\ra, \eqn{Ti_conv}
\ee
or, without the spectator $\delta$ function, as
\be
\tG_0 \tilde{T}_i\tG_0 =  \la D_{0i} t_i D_{0i}\ra  \otimes \la d_i\ra \eqn{tmp}
\ee 
 where the symbol $\otimes$ denotes the convolution integral:
\be
a\otimes b (E) \equiv  \frac{i}{2\pi} \int_{-\infty}^\infty a(E-z) b(z)\, dz.
\ee

The above analysis of three-body Green functions can be repeated for two-body Green functions, thereby yielding the two-body version of \eq{main}:
\be
 \tD_0 \tilde{t} \tD_0 = \la D_0 t D_0\ra,   \eqn{main_2b}
\ee
where
\be
\tilde{t} = \tv +\frac{1}{2}\tv \tD_0 \tilde{t}.   \eqn{tT_LS}
\ee
Using this in \eq{tmp} we obtain the essential result
\be
 \tG_0 \, \tilde{T}_i \, \tG_0  =   \tD_{0i} \tilde{t}_i \tD_{0i}\otimes\la d_i\ra
 \eqn{conv}
\ee
which expresses the exact t matrix $\tilde{T}_i$, forming the 'spectator plus interacting pair' input to the three-body Faddeev equations,  in terms of a convolution of the spectator propagator and the sub-system equal-time two-body t matrix.

\section{In-matter three-dimensional three-body equations}  

\subsection{General description}
The unsymmetrized three-body equal time Green function $\tG^u$ is given in terms of the three-body t matrix $\tT$ by \eq{tGuT}. If the three-body force $\tK_c$ of \eq{tV} is neglected, one can express  $\tT$ in the Faddeev form
\be
\tT = \sum_{i=1}^3 X_i 
 \ee
where
\be
X_i =\frac{1}{2} \tT_i +\frac{1}{2} \tT_i \sum_{k\ne i} \tG_0 X_k .  \eqn{Faddeev}
\ee
Alternatively, one can express $\tG^u$ in terms of Alt-Grassberger-Sandhas (AGS) amplitudes $U_{ij}$ \cite{AGS},
\be
\tG^u = \tG_i \delta_{ij} + \tG_i U_{ij} \tG_j
\ee 
where the $U_{ij}$ satisfy the AGS equations
\be
U_{ij} = \tG_0^{-1}\bar{\delta}_{ij} +\frac{1}{2} \sum_{k} \bar{\delta}_{ik}\tT_k \tG_0 U_{kj}.   \eqn{AGS}
\ee
In either case, the input consists of  the disconnected amplitudes $\tT_i$ which are specified in terms of equal-time two-body t matrices $\tilde{t}_i$ according to \eq{conv}.

The above equations constitute our general formulation of the finite temperature equal-time in-matter three-body problem. What is noteworthy is that the neglect of the three-body force $\tK_c$ is the only approximation made; in particular, the four-dimensional two-body potential $V_i$ of \eq{V_i_sym}, which is specified by the underlying Hamiltonian,  is included exactly and to all orders within the equal-time approach. Although the above equations can be used directly for calculations, in the next subsection we shall show that for the case of instantaneous potentials and effective single-particle dressings, they can be greatly simplified. 

\subsection{Description for instantaneous potentials and free-like dressed propagators}
Here we consider the commonly used approximations where the four-dimensional two-body potential $v$ is assumed to be instantaneous, and where single particle dressings are taken into account only through effective masses and effective chemical potentials.

Thus we assume that the dressed propagator $d$ has exactly the same structure as the free propagator $d^f$, given in \eq{df}, but with a modified (effective) mass $m^*$ and a modified chemical potential $\mu^*$. From \eq{df_tt} this means we can write for particle $i$,
\begin{align}
\la d_i\ra  &= \bn_i d_i^r  + n_i d_i^a  \eqn{d_ira}\\
&\equiv \la d^r_i\ra + \la d^a_i\ra
\end{align}
where
\be
d_i^{r,a} = \frac{1}{E-\w_i \pm  i\epsilon},
\ee
$d_i^r$ being specified with $+i\epsilon$ and $d_i^a$ with $-i\epsilon$, $\bn_i=\bn(\p_i)$, $n_i=n(\p_i)$, and $\w_i = \p_i^2/2m^*-\mu^*$.
The disconnected equal-time two-body propagator, $\la D_0\ra$, will then take the same form as $\la D_0^f\ra$ given in  \eq{D0f}, and we similarly write
\begin{align}
\la D_{0i}\ra &=\bn_j\bn_kD^r_i- n_j n_kD^a_i\\
&\equiv  \la D^r_{0i}\ra + \la D^a_{0i}\ra \eqn{D0i}
\end{align}
where
\begin{align}
D^{r,a}_i&=\frac{1}{E-\w_j-\w_k\pm i\epsilon}  . \eqn{Dra}
\end{align}
The disconnected equal-time three-body propagator, $\la G_0\ra$, can likewise be written as
\begin{align}
\la G_{0}\ra &=\bn_i\bn_j\bn_kG^r+ n_i n_j n_kG^a\\
&\equiv  \la G^r_{0}\ra + \la G^a_{0}\ra
\end{align}
where
\begin{align}
G^{r,a}&=\frac{1}{E-\w_i-\w_j-\w_k\pm i\epsilon}  . \eqn{Gra}
\end{align}

At the same time, the assumption of an instantaneous two-body potential means, in momentum space, that potential $v$, and therefore the corresponding t matrix $t$, do not depend of zero components of relative momenta. Thus $\la D_0 t D_0\ra = \la D_0\ra it \la D_0\ra$, so that \eq{main_2b} becomes
\be
\tD_0 \tilde{t} \tD_0 =  \la D_0\ra it \la D_0\ra.   \eqn{equal_t1}
\ee
Now, because the disconnected equal-time two-body propagator $\la D_0\ra$ has the form specified by \eq{D0i}, it follows that  $\la D_0\ra = \la D_0\ra gg {\cal N}^{(2)}$, and therefore $\la D_0\ra = \tD_0 gg {\cal N}^{(2)}$. Applying this to \eq{equal_t1}, one obtains the useful identity
\be
\la D_0\ra \tilde{t} \la D_0\ra = \la D_0\ra it \la D_0\ra = \tD_0\tilde{t}\tD_0.   \eqn{equal_t}
\ee
Interestingly, the quasi-potential $\tv$ and the instantaneous potential $v$ obey a similar equation,
\be
\la D_0\ra \tilde{v} \la D_0\ra = \la D_0\ra iv \la D_0\ra = \tD_0\tilde{v}\tD_0.   \eqn{equal_v}
\ee
Indeed, for the case where $v$ is instantaneous and $\la D_0\ra$ is specified by \eq{D0i}, 
 \eq{main_2b}  holds also for potentials, i.e.,
\be
 \tD_0 \tilde{v} \tD_0 = \la D_0 v D_0\ra.    \eqn{main_v}
\ee
To see this, consider the second order term in $v$ of \eq{tvpert} under the assumption of instantaneous $v$. Up to a constant factor, we have that
\begin{align}
\la D_0 v &D_0 v D_0\ra - \la D_0 v D_0\ra \tD_0^{-1}\la D_0 v D_0\ra\nn
&= \la D_0\ra iv  \la D_0\ra iv  \la D_0\ra -  \la D_0\ra iv  \la D_0\ra \tD_0^{-1} \la D_0\ra iv  \la D_0\ra\nn
&=  \la D_0\ra iv  \la D_0\ra \left( 1 -   \tD_0^{-1} \la D_0\ra\right)  iv  \la D_0\ra.
\end{align}
Then taking $\la D_0\ra$ to be given as in \eq{D0i}, so that $\la D_0\ra = \la D_0\ra gg {\cal N}^{(2)}=  \tD_0 gg {\cal N}^{(2)}$, 
\begin{align}
\la D_0 v &D_0 v D_0\ra - \la D_0 v D_0\ra \tD_0^{-1}\la D_0 v D_0\ra\nn
&=  \la D_0\ra iv  \la D_0\ra \left( gg {\cal N}^{(2)} -   \tD_0^{-1} \la D_0\ra\right)  iv  \la D_0\ra\\
&=  \la D_0\ra iv  \la D_0\ra\tD_0^{-1} \left( \tD_0\, gg\, {\cal N}^{(2)} -    \la D_0\ra\right)  iv  \la D_0\ra = 0.
\end{align}
In a similar way, all higher order contributions in $v$ are zero in \eq{tvpert}, and the result of \eq{main_v}, follows.

Applying \eq{equal_t} to \eq{conv}, one obtains
\be
 \tG_0 \, \tilde{T}_i \, \tG_0  =   \la D_{0i} \ra\tilde{t}_i \la D_{0i}\ra \otimes\la d_i\ra   \eqn{B4}
\ee
which is to be used as the input to the equal-time Faddeev equations discussed above.

\subsubsection{Split form of equations}
Under the assumptions of this subsection, the disconnected two- and three-particle dressed propagators, $\la D_0\ra$ and $\la G_0\ra$, have the same projection properties as the corresponding free propagators. These projection properties lead to a substantial simplification of the in-matter three-body equations.  We shall show that for instantaneous potentials and the free-like dressed propagators of \eq{Dra} and \eq{Gra},  the in-matter Faddeev equation, \eq{Faddeev}, can be split into two equations, one involving only retarded parts of propagators, the other involving only the advanced parts. 

In view of \eq{B4}, we begin by defining the quantities $\tT^R_i$ and $\tT^A_i$  by\footnote{The superscripts of $\tT_i^{R,A}$ are simply convenient labels and do not correspond exactly to the retarded or advanced $\theta$-functions of time as, for example, in the case of the single particle propagator of \eq{d_ira}.}
\begin{subequations}
\begin{align}
\tG_0\, \tT^R_i \, \tG_0 &= (2\pi)^3\delta^3(\p_i'-\p_i)\,\la D_{0i}\ra \tilde{t}_i \la D_{0i} \ra 
\otimes \la d^r_i\ra,  \eqn{TR} \\
\tG_0\, \tT^A_i \, \tG_0 &= (2\pi)^3\delta^3(\p_i'-\p_i)\,\la D_{0i}\ra \tilde{t}_i \la D_{0i} \ra 
\otimes \la d^a_i\ra,  \eqn{TA}
\end{align}
\end{subequations}
so that
\be
\tT_i = \tT^R_i + \tT^A_i.
\ee
One can now check that $\tT^R_i \tG_0 \tT^A_k=0$, when $i\neq k$:
\begin{align}
\tG_0 \tT^R_i \tG_0 \tT^A_k \tG_0 & =
\la D_{0i} \ra \tilde{t}_i \la D_{0i}\ra \otimes \la d^{r}_i\ra \, \tG_0^{-1}\,
\la D_{0k}\ra \tilde{t}_k\la D_{0k}\ra \otimes \la d^{a}_k\ra \nn
&= 
\la D_{0i} \ra \tilde{t}_i 
(\bn_j\bn_kD^r_i- n_j n_kD^a_i)\otimes  \bn_i d^r_i\, \tG_0^{-1} \nn
&\hspace{5mm}
(\bn_i\bn_jD^r_k- n_i n_jD^a_k)\tilde{t}_k\la D_{0k}\ra \otimes n_k d^{a}_k
=0,
\end{align}
where we used \eq{tG0_inv} for $\tG_0^{-1}$, and the fact that
\be
(\bn_i\bn_j\bn_kD^r_i- \bn_in_j n_kD^a_i) ggg (\bn_i\bn_jn_kD^r_k- n_i n_jn_kD^a_k) = 0.
\ee
It is therefore clear that the solution to the Faddeev equation, \eq{Faddeev},
has the form $X_i=X^R_i+X^A_i$, where $X^R_i$ and $X^A_i$ satisfy the
independent equations
\begin{subequations}
\begin{align}
X^R_i &=\frac{1}{2}\tT^R_i+\frac{1}{2}\tT^R_i\sum_{k\neq i}\tG_0 X^R_k,\\
X^A_i &=\frac{1}{2}\tT^A_i+\frac{1}{2}\tT^A_i\sum_{k\neq i}\tG_0 X^A_k.
\end{align}\eqn{Aspliteq}
\end{subequations}
In a similar way we have that
\begin{align}
\tG_0\tT^R_i \tG_0 \tT^R_k\tG_0 & = 
\la D_{0i} \ra \tilde{t}_i \la D_{0i}\ra \otimes \la d^{r}_i\ra \, \tG_0^{-1}\,
\la D_{0k}\ra \tilde{t}_k\la D_{0k}\ra \otimes \la d^{r}_k\ra \nn
&= 
\la D_{0i} \ra \tilde{t}_i 
(\bn_j\bn_kD^r_i- n_j n_kD^a_i)\otimes  \bn_i d^r_i\, \tG_0^{-1} \nn
&\hspace{6mm}
(\bn_i\bn_jD^r_k- n_i n_jD^a_k)\tilde{t}_k\la D_{0k}\ra \otimes \bn_k d^{r}_k\nn
&=
\la D_{0i} \ra \tilde{t}_i 
(\bn_j\bn_kD^r_i)\otimes  \bn_i d^r_i\, \tG_0^{-1}
(\bn_i\bn_jD^r_k)\tilde{t}_k\la D_{0k}\ra \otimes \bn_k d^{r}_k\nn
&=
\la D_{0i} \ra \tilde{t}_i \la D^r_{0i}\ra \otimes \la d^{r}_i\ra \, \tG_0^{-1}\,
\la D^r_{0k}\ra \tilde{t}_k\la D_{0k}\ra \otimes \la d^{r}_k\ra,  \eqn{RR}
\end{align}
and
\begin{align}
\tG_0\tT^A_i \tG_0 \tT^A_k \tG_0& =
\la D_{0i} \ra \tilde{t}_i \la D_{0i}\ra \otimes \la d^a_i\ra \, \tG_0^{-1}\,
\la D_{0k}\ra \tilde{t}_k\la D_{0k}\ra \otimes \la d^a_k\ra \nn
&= 
\la D_{0i} \ra \tilde{t}_i 
(\bn_j\bn_kD^r_i- n_j n_kD^a_i)\otimes  n_i d^a_i\, \tG_0^{-1} \nn
&\hspace{6mm}
(\bn_i\bn_jD^r_k- n_i n_jD^a_k)\tilde{t}_k\la D_{0k}\ra \otimes n_k d^a_k\nn
&=
\la D_{0i} \ra \tilde{t}_i 
(-n_jn_kD^a_i)\otimes  \bn_i d^r_i\, \tG_0^{-1}
(-n_in_jD^a_k)\tilde{t}_k\la D_{0k}\ra \otimes \bn_k d^a_k\nn
&=
\la D_{0i} \ra \tilde{t}_i \la D^a_{0i}\ra \otimes \la d^a_i\ra \, \tG_0^{-1}\,
\la D^a_{0k}\ra \tilde{t}_k\la D_{0k}\ra \otimes \la d^a_k\ra.  \eqn{AA}
\end{align}
Thus, apart from external two-body propagators $\la D_{0i} \ra$ and $\la D_{0k} \ra$, all other propagators explicitly shown on the RHS of \eqs{RR} and (\ref{AA})  are truncated to their retarded or advanced parts. To take advantage of this simplification, we iterate \eqs{Aspliteq} once to obtain
\begin{subequations}
\begin{align}
X_i^R &= \frac{1}{2}\tT_i^{R} + \frac{1}{4}\sum_{k\ne i}\tT_i^{Rr}(\tG_0 + \sum_{j\ne k} \tG_0 X_j^r\tG_0)\tT_k^{\,r\!R}\\
X_i^A &= \frac{1}{2}\tT_i^{A} + \frac{1}{4}\sum_{k\ne i}\tT_i^{Aa}(\tG_0 + \sum_{j\ne k}\tG_0 X_j^a\tG_0)\tT_k^{\,aA}
\end{align}  \eqn{XRA1}
\end{subequations}
where
\begin{subequations}
\begin{align}
\tG_0\, \tT^{Rr}_i \, \tG_0 &= (2\pi)^3\delta^3(\p_i'-\p_i)\,\la D_{0i}\ra \tilde{t}_i \la D^r_{0i} \ra 
\otimes \la d^r_i\ra, \\
\tG_0\, \tT^{\,r\!R}_i \, \tG_0 &= (2\pi)^3\delta^3(\p_i'-\p_i)\,\la D^r_{0i}\ra \tilde{t}_i \la D_{0i} \ra 
\otimes \la d^r_i\ra,\\
\tG_0\, \tT^{Aa}_i \, \tG_0 &= (2\pi)^3\delta^3(\p_i'-\p_i)\,\la D_{0i}\ra \tilde{t}_i \la D^a_{0i} \ra 
\otimes \la d^a_i\ra, \\
\tG_0\, \tT^{\,a A}_i \, \tG_0 &= (2\pi)^3\delta^3(\p_i'-\p_i)\,\la D^a_{0i}\ra \tilde{t}_i \la D_{0i} \ra 
\otimes \la d^a_i\ra,
\end{align}   \eqn{Ts}
\end{subequations}
and where the amplitudes $X_i^r$ and $X_i^a$ satisfy the Faddeev equations
\begin{subequations}
\begin{align}
X^r_i &=\frac{1}{2}\tT^r_i+\frac{1}{2}\tT^r_i\sum_{k\neq i}\tG_0 X^r_k, \eqn{Xr1}\\
X^a_i &=\frac{1}{2}\tT^a_i+\frac{1}{2}\tT^a_i\sum_{k\neq i}\tG_0 X^a_k, \eqn{Xa1}
\end{align}\eqn{Xra1}
\end{subequations}
which are simpler than \eqs{Aspliteq} in that they utilize input t matrices whose adjoining propagators are either all retarded or all advanced:
\begin{subequations}
\begin{align}
\tG_0\, \tT^{r}_i \, \tG_0 &= (2\pi)^3\delta^3(\p_i'-\p_i)\,\la D^r_{0i}\ra \tilde{t}_i \la D^r_{0i} \ra 
\otimes \la d^r_i\ra, \eqn{Tr1}\\
\tG_0\, \tT^{a}_i \, \tG_0 &= (2\pi)^3\delta^3(\p_i'-\p_i)\,\la D^a_{0i}\ra \tilde{t}_i \la D^a_{0i} \ra 
\otimes \la d^a_i\ra. \eqn{Ta1}
\end{align} \eqn{T1}
\end{subequations}
Note, however, that 
\be
\tilde{t}_i = \tv_i + \frac{1}{2}\tv_i \la D_{0i}\ra \tilde{t}_i, \eqn{tt_LS}
\ee
so that the internal propagators, $\la D_{0i}\ra$, used in constructing the physical two-body t matrices $\tilde{t}_i$, retain both retarded and advanced parts. With the spectator $\delta$ function removed,
one can invert the $\tG_0$'s in \eqs{T1} with the help of \eq{tG0_inv}, to obtain
\begin{subequations}
\begin{align}
\tilde{T}_i^r(E)  &= \frac{i}{2\pi}\int_{-\infty}^\infty dz\, \frac{ggg\, \bn_j'\bn_k'\,(E-\w_{ijk}')\,
\tilde{t}_i(E-z)\,(E-\w_{ijk})\,\bn_i\bn_j\bn_k \,ggg}{(E-z-\w_{jk}'+i\epsilon)
(E-z-\w_{jk}+i\epsilon)(z-\w_i+i\epsilon)}\\
\tilde{T}_i^a(E)  &= \frac{i}{2\pi}\int_{-\infty}^\infty dz\, \frac{ggg\, n_j'n_k'(E-\w_{ijk}')\,
\tilde{t}_i(E-z)\,(E-\w_{ijk})\,n_in_jn_k\,ggg}{(E-z-\w_{jk}'-i\epsilon)
(E-z-\w_{jk}-i\epsilon)(z-\w_i-i\epsilon)}
\end{align} \eqn{T3}
\end{subequations}
where $\w_{ijk}'=\w_i+\w_j'+\w_k'$, $\w_{ijk}=\w_i+\w_j+\w_k$, $\w_{jk}'=\w_j'+\w_k'$, and
$\w_{jk}=\w_j+\w_k$. In turn, this result shows that the $\tG_0$ in \eq{Xr1} can be replaced by $\la G_0^r\ra$, and the 
$\tG_0$ in \eq{Xa1} can be replaced by $\la G_0^a\ra$. Similarly, replacements can be made in appropriate $\tG_0$'s in \eq {XRA1} and \eq{Ts}, so that our final equations for the amplitude $X_i$ are:
\be
X_i = X_i^R + X_i^A,   \eqn{X_i}
\ee
\begin{subequations}
\begin{align}
X_i^R &= \frac{1}{2}\tT_i^{R} + \frac{1}{4}\sum_{k\ne i}\tT_i^{Rr}\left(\la G_0^r\ra + \sum_{j\ne k}
\la G_0^r\ra X_j^r \la G_0^r\ra \right)\tT_k^{\,r\!R} ,  \eqn{XR} \\
X_i^A &= \frac{1}{2}\tT_i^{A} + \frac{1}{4}\sum_{k\ne i}\tT_i^{Aa}\left(\la G_0^a\ra + \sum_{j\ne k}
\la G^a_0\ra X_j^a\la G^a_0\ra\right)\tT_k^{\,aA}, \eqn{XA}
\end{align}  \eqn{XRA}
\end{subequations}
where the amplitudes $X_i^r$ and $X_i^a$ satisfy the Faddeev equations
\begin{subequations}
\begin{align}
X^r_i&=\frac{1}{2}\tT^r_i+\frac{1}{2}\tT^r_i\sum_{k\neq i}\la G^r_0\ra X^r_k,
\eqn{Xr}\\
X^a_i&=\frac{1}{2}\tT^a_i+\frac{1}{2}\tT^a_i\sum_{k\neq i}\la G^a_0\ra X^a_k.
\eqn{Xa}
\end{align}  \eqn{Xra}
\end{subequations}

\subsubsection{Single degree of freedom equations}

The projection properties of the input two-body t matrices of \eqs{T3} enable
one to eliminate the doubled degrees of freedom from the in-matter Faddeev
equations, \eq{Xra}. To see this, we first note that the amplitude $X_j^R$ of
\eqs{XR} is determined by the quantity
\be
\la G^r_0\ra X_j^r\la G^r_0\ra = G^r \bn_1\bn_2\bn_3 X_j^r \bn_1\bn_2\bn_3 G^r, 
\ee
which has the doubled-degrees of freedom Faddeev amplitude $X_j^r$ completely surrounded by projection operators $\bn$.
The matrix structure of $\bn$ is given in \eq{bn} as
\be
\bn = \uu(\w) \uu^\dagger(\w)
\ee
where $\uu(\w)$ is a column vector defined in \eq{uvB} for bosons and in \eq{uvF} for
fermions, and $\uu^\dagger(\w)$ is the corresponding row vector. Thus amplitude $X_j^R$ is expressible directly in terms of the single-degree of freedom amplitude
\be
\hX_j^r =
\uu^\dagger(\omega_1')\uu^\dagger(\omega_2')\uu^\dagger(\omega_3')\,
X_j^r\, \uu(\omega_1)\uu(\omega_2)\uu(\omega_3). \eqn{hXr}
\ee
Similarly, the matrix structure of projection operator $n$, given in \eq{n} as
\be
n = \pm \vv(\w) \vv^\dagger(\w),
\ee
allows one to express amplitude $X_j^A$ of \eq{XA} directly in terms of the single-degree of freedom amplitude
\be
\hX_j^a =
\vv^\dagger(\omega_1')\vv^\dagger(\omega_2')\vv^\dagger(\omega_3')\,
X_j^a\, \vv(\omega_1)\vv(\omega_2)\vv(\omega_3). \eqn{hXa}
\ee
Moreover, it follows from \eqs{Xra} that $\hX_i^r$ and $\hX_i^a$ themselves
satisfy Faddeev equations
\begin{subequations}
\begin{align}
\hX_i^r &= \frac{1}{2}\hat{T}^r_i +\frac{1}{2}\hat{T}^r_i\sum_{k\ne i} G^r
\hX_k^r \\
\hX_i^a &= \frac{1}{2}\hat{T}^a_i +\frac{1}{2}\hat{T}^a_i\sum_{k\ne i} G^a
\hX_k^a
\end{align}  \eqn{hX}
\end{subequations}
where, with the spectator $\delta$ function removed,
\begin{subequations}
\begin{align}
\hat{T}_i^r(E)  &= \frac{i}{2\pi}\int_{-\infty}^\infty dz\,
\frac{(E-\w_{ijk}')\,
\hat{t}^R_i(E-z)\,(E-\w_{ijk})}{(E-z-\w_{jk}'+i\epsilon)
(E-z-\w_{jk}+i\epsilon)(z-\w_i+i\epsilon)}\\
\hat{T}_i^a(E)  &= \frac{i}{2\pi}\int_{-\infty}^\infty dz\,
\frac{(E-\w_{ijk}')\,
\hat{t}^A_i(E-z)\,(E-\w_{ijk})}{(E-z-\w_{jk}'-i\epsilon)
(E-z-\w_{jk}-i\epsilon)(z-\w_i-i\epsilon)}
\end{align} \eqn{T4}
\end{subequations}
and
\begin{subequations}
\begin{align}
\hat{t}^R_i & = \uu^\dagger(\omega_j')\uu^\dagger(\omega_k')\,
\tilde{t}_i\, \uu(\omega_j)\uu(\omega_k),\\[2mm]
\hat{t}^A_i & = \vv^\dagger(\omega_j')\vv^\dagger(\omega_k')\,
\tilde{t}_i\, \vv(\omega_j)\vv(\omega_k).
\end{align} \eqn{ht}
\end{subequations}
We note that in \eqs{hX} all doubled-degrees of freedom have been
eliminated. In particular, the two-body input to these equations is found from
the convolution integrals, \eqs{T4}, involving the single-degree of freedom t
matrices $\hat{t}^R_i$ and $\hat{t}^A_i$ of \eqs{ht}.

Although both $\hat{t}^R_i$ and $\hat{t}^A_i$ are single-degree of freedom
quantities, they themselves are constructed from a 16-component doubled-degree
of freedom t matrix $\tilde{t}_i$. Although it may not be difficult to solve
the 16-component equation, \eq{tt_LS}, to obtain $\tilde{t}_i$, it is useful to
note that this equation can be recast into a four-component ($2\times 2$) equation as
follows. Writing
\eq{tt_LS} as
\be
\tilde{t}_i = \tilde{v}_i + \frac{1}{2} \tilde{v}_i \left( \bn_j\bn_k D_i^r
- n_j n_k D_i^a\right)\tilde{t}_i,
\ee
straightforward use of \eqs{bn} and (\ref{n}) allows one to write
\be
\hat{t}_i = \hat{v}_i + \frac{1}{2} \hat{v}_i \hat{D}_i \hat{t}_i,
\ee
where
\begin{align}
\hat{t}_i &= \left(\begin{array}{cc} \ud_j\ud_k \tilde{t}_i \uu_j \uu_k &
\ud_j\ud_k \tilde{t}_i \vv_j \vv_k\\[2mm]
\vd_j\vd_k \tilde{t}_i \uu_j \uu_k &\vd_j\vd_k \tilde{t}_i \vv_j \vv_k
\end{array}\right),\\[3mm]
\hat{v}_i & = \left(\begin{array}{cc} \ud_j\ud_k \tilde{v}_i \uu_j \uu_k &
\ud_j\ud_k \tilde{v}_i \vv_j \vv_k\\[2mm]
\vd_j\vd_k \tilde{v}_i \uu_j \uu_k &\vd_j\vd_k \tilde{v}_i \vv_j \vv_k
\end{array}\right),\\[2mm]
\hat{D}_i &= \left(\begin{array}{cc} D_i^r & 0 \\
0 & - D_i^a
\end{array}\right).
\end{align}

\section{Summary}

Using the real-time formalism, we have formulated equal-time three-body equations that describe three identical particles interacting via pair-wise interactions at finite temperature and density. 
Starting with the four-dimensional field-theoretic description of the $3\rightarrow 3 $ Green function, equal-time three-body equations were derived without resorting to any approximations beyond that of the assumption of pair-wise interactions. 

Our resulting in-matter three-body equations, \eq{Faddeev} for the general case, and \eq{X_i}-\eq{Xra} for the case of instantaneaous potentials and free-like dressed propagators, have the familiar Faddeev form, although they differ from the usual zero-density Faddeev equations in that they involve doubled degrees of freedom (inherent in the real-time formalism), and they utilize one-body thermal Green functions (which have retarded and advanced parts, and depend on both temperature and chemical potential). At the same time, the form of our equations is similar to that of other formulations of the in-matter three-body problem, even though all other formulations have apparently been done either at zero temperature, or for non-zero temperatures, using the imaginary time formalism. 
However, what distinguishes our approach in an essential way from all other derivations of the equal-time in-matter three-body problem, is that we have managed to avoid any approximations in the equal-time 3D reduction of the original 4D field-theoretic formulation. Moreover, our resulting 3D equations remain practical in that the equal-time two-body t matrix in three-body space, $\tT_i$, which determines the integral equation kernel, is given in terms of the 4D two-body t matrix by a simple convolution integral - \eq{tmp}.

\begin{acknowledgments}
The research described in this publication was made possible in part by Award No.\ GEP2-3329-TB-03 ofÊ the Georgian Research and Development Foundation (GRDF) and the U.S. Civilian Research \& Development Foundation for the Independent States of the Former Soviet Union (CRDF).
\end{acknowledgments}

\appendix
\section{Propagator matrix structure}

In the doubled degrees of freedom formalism, the matrix structure of propagators is discussed in detail, for example, in Ref.\ \cite{umezawa}. Here we give only a brief summary.

\subsection{Bosons}

In the doubled degrees of freedom formalism, the boson propagator is given by \cite{umezawa}
\be
d^f(p)=i\mat{\cosh\w}{\sinh\w}{\sinh\w}{\cosh\w} \mat{d^r}{0}{0}{-d^a}
\mat{\cosh\w}{\sinh\w}{\sinh\w}{\cosh\w}    \eqn{doubprop}
\ee
where
\be
d^r = \frac{1}{p^0-\w+i\epsilon}, \hspace{7mm} d^a = \frac{1}{p^0-\w-i\epsilon}, \hspace{7mm}
\w=\frac{\p^2}{2m}-\mu  \eqn{dw}
\ee
and
\be
\sinh\w = \sqrt{f_B(\w)} ,\hspace{7mm}\cosh\w = \sqrt{1 +f_B(\w)} ,\hspace{7mm}
f_B(\w)= \frac{1}{e^{\beta\w}-1}.
\ee
Denoting
\be
U_B(\w)=\mat{\cosh\w}{\sinh\w}{\sinh\w}{\cosh\w},\hspace{1cm}
g_B=\mat{1}{0}{0}{-1}, \eqn{UB}
\ee
one finds that $U_B$ is not unitary but satisfies
\be
U_B(\w) \,g_B\, U^\dagger_B(\w) = g_B.
\ee
It follows that
\be
\left[d^f(p)\right]^{-1} = -i(p^0-\w) g_B.
\ee
Comparison with \eq{df} provides explicit expressions for $\bn$ and $n$:
\be
\bn = U_B(\w) \mat{1}{0}{0}{0} U^\dagger_B(\w),\hspace{7mm}
 n = U_B(\w) \mat{0}{0}{0}{-1} U^\dagger_B(\w).  \eqn{nB}
\ee
One can check the following important properties of $n$ and $\bn$:
\begin{subequations}
\begin{align}
n + \bn &= g_B\\
n\, g_B\, \bn &= \bn\, g_B\, n = 0\\
n \,g_B \,n &= n,\hspace{3mm}
\bn\, g_B\,\bn = \bn .
\end{align}  \eqn{bproj}
\end{subequations}
For the case of bosons, these relations will define what is meant by the projection properties of 
the operators $n$ and $\bn$.

\eq{nB} implies another convenient way of expressing $\bn$ and $n$, namely, as products of
column and row vectors:
\be
\bn = \uu_B(\w) \uu_B^\dagger(\w), \hspace{1cm} n = - \vv_B(\w) \vv_B^\dagger(\w)
\ee
where
\begin{align}
\uu_B(\w) & = \left(\begin{array}{c} U^B_{11}\\[2mm] U^B_{21}\end{array}\right)
=\left(\begin{array}{c} \cosh\w \\[2mm] \sinh\w\end{array}\right),\hspace{5mm}
\vv_B(\w) = \left(\begin{array}{c} U^B_{12}\\[2mm] U^B_{22}\end{array}\right)
=\left(\begin{array}{c} \sinh\w \\[2mm] \cosh\w\end{array}\right), \eqn{uvB}
\end{align}
are the column vectors and $\uu_B^\dagger$, $\vv_B^\dagger$, are the corresponding row vectors.
It follows that
\be
\uu^\dagger_B\, g_B\, \uu_B = 1,\hspace{1cm} \vv^\dagger_B\, g_B\, \vv_B = -1.
\ee
\subsection{Fermions}

In the doubled degrees of freedom formalism, the fermion propagator is given by \cite{umezawa}
\be
d^f(p) = i \mat{\cos\w}{\sin\w}{-\sin\w}{\cos\w} \mat{d^r}{0}{0}{d^a}
\mat{\cos\w}{-\sin\w}{\sin\w}{\cos\w}
\ee
where $d^r$, $d^a$, $\omega$ are as in \eq{dw}, and
\be
\sin\w=\sqrt{f_F(\w)},\hspace{7mm}
\cos\w=\sqrt{1-f_F(\w)},\hspace{7mm}f_F(\w)=
\frac{1}{e^{\beta\w}+1}.
\ee
Denoting
\be
U_F(\w)=\mat{\cos\w}{\sin\w}{-\sin\w}{\cos\w},  \eqn{UF}
\ee
it's clear that $U_F$ is unitary:
\be
U_F(\w) U^\dagger_F(\w) = U^\dagger_F(\w) U_F(\w) = 1.
\ee
It follows that
\be
\left[d^f(p)\right]^{-1} = -i(p^0-\w) \mat{1}{0}{0}{1}.
\ee
Comparison with \eq{df} provides explicit expressions for $\bn$ and $n$:
\be
\bn = U_F(\w) \mat{1}{0}{0}{0} U^\dagger_F(\w),\hspace{7mm}
 n = U_F(\w) \mat{0}{0}{0}{1} U^\dagger_F(\w).
\ee
One can check the following projection properties of $n$ and $\bn$:
\begin{subequations}
\begin{align}
n + \bn &= 1\\
n\, \bn &= \bn\, n = 0\\
n \,n &= n,\hspace{3mm}
\bn\,\bn = \bn .
\end{align}  \eqn{fproj}
\end{subequations}
As in the boson case, we can write $\bn$ and $n$ in terms of column and row
vectors:
\be
\bn = \uu_F(\w) \uu_F^\dagger(\w), \hspace{1cm} n = \vv_F(\w) \vv_F^\dagger(\w)
\ee
where
\begin{align}
\uu_F(\w) & = \left(\begin{array}{c} U^F_{11}\\[2mm] U^F_{21}\end{array}\right)
=\left(\begin{array}{c} \cos\w \\[2mm] -\sin\w\end{array}\right),\hspace{5mm}
\vv_F(\w) = \left(\begin{array}{c} U^F_{12}\\[2mm] U^F_{22}\end{array}\right)
=\left(\begin{array}{c} \sin\w \\[2mm] \cos\w\end{array}\right), \eqn{uvF}
\end{align}
are the column vectors and $\uu_F^\dagger$, $\vv_F^\dagger$, are the corresponding row vectors.
It follows that
\be
\uu^\dagger_F\, \uu_F = 1,\hspace{1cm} \vv^\dagger_F\, \vv_F = 1.
\ee

\subsection{General}
By defining
\be
g_F = \mat{1}{0}{0}{1}, \hspace{5mm}g_B = \mat{1}{0}{0}{-1}, \hspace{5mm}
\ee
and
\be
g = \left\{\begin{array}{l} g_B \\[1mm]
 g_F
 \end{array}\right. , \hspace{5mm}
U = \left\{\begin{array}{l} U_B \\[1mm]
 U_F
\end{array}\right. ,\hspace{5mm}
\uu = \left\{\begin{array}{l} \uu_B\\[1mm]
 \uu_F
\end{array}\right. ,\hspace{5mm}
\vv = \left\{\begin{array}{ll} \vv_B & \hspace{5mm}(\mbox{boson case})\\[1mm]
 \vv_F & \hspace{5mm}(\mbox{fermion case})
\end{array}\right.
\ee
one can write, for both bosons and fermions,
\begin{align}
d^f(p) &=i \, U g \mat{d^r}{0}{0}{d^a} U^\dagger \\[2mm]
\left[d^f(p)\right]&^{-1} = -i (p^0-\w) g,
\end{align}
\begin{align}
U &\,g\, U^\dagger = U^\dagger\, g\, U = g, \\[4mm]
\bn &= U  \mat{1}{0}{0}{0} U^\dagger,\hspace{7mm}
 n = U g  \mat{0}{0}{0}{1} U^\dagger,  \eqn{U}
\end{align}
\begin{subequations}
\begin{align}
n + \bn &= g\\
n\, g\, \bn &= \bn\, g\, n = 0\\
n \,g \,n &= n,\hspace{3mm}
\bn\, g\,\bn = \bn .
\end{align}  \eqn{proj}
\end{subequations}
Also, for both bosons and fermions,
\be
\bn = \uu \uu^\dagger,\hspace{1cm} \uu^\dagger\, g\, \uu = 1,  \eqn{bn}
\ee
while
\be
n = \pm \vv \vv^\dagger,\hspace{1cm} \vv^\dagger\, g\, \vv = \pm 1,  \eqn{n}
\ee
with the $+$ sign being for fermions and $-$ sign being for bosons.

\end{document}